\documentclass[twocolumn]{jpsj2}

\usepackage{bm}

\renewcommand{\d}{\text{d}}

\title{%
Molecular Chemical Engines: Pseudo-Static Processes
and the Mechanism of Energy Transduction}

\author{%
Kazuo \textsc{Sasaki}\thanks{
E-mail address: sasaki@camp.apph.tohoku.ac.jp.} 
}

\inst{%
Department of Applied Physics, Tohoku University, 
6-6-05 Aoba-yama, Sendai 980-8579
}

\recdate{\today}

\abst{%
We propose a simple theoretical model for a molecular
chemical engine that catalyzes a chemical reaction
and converts the free energy released 
by the reaction into mechanical work.
Binding and unbinding processes of reactant and product molecules to and from the engine are explicitly taken into account.
The work delivered by the engine is calculated analytically
for infinitely slow (``pseudo-static'') processes, 
which can be reversible (quasi-static) or irreversible,
controlled by an external agent.
It is shown that the work larger than the maximum value
limited by the second law of thermodynamics can be
obtained in a single cycle of operation by chance,
although the statistical average of the work
never exceeds this limit and the maximum work is delivered 
if the process is reversible.
The mechanism of the energy transduction 
is also discussed.
}

\kword{%
molecular chemical engine, quasi-static process, the second law of thermodynamics, energy transduction}

\begin{document}

\sloppy
\maketitle

\section{\label{sec:intro}Introduction}

Biological molecular motors, such as myosin, kinesin, dynein and 
$\rm F_1$-ATPase, are tiny machines that catalyze 
the hydrolysis reaction of adenosine triphosphate (ATP)
and convert the free energy released by the reaction into mechanical work.%
\cite{alberts02,schliwa03,howard01}
The mechanisms of energy transduction in these \textit{molecular chemical 
engines} are not fully understood.
A simple, idealized model of a molecular chemical engine,
like the Carnot cycle of a heat engine, would provide clear understanding
of the energy transduction in biomolecular motors
and insights into the design of artificial nanomachines.
The purpose of the present paper is to propose such a model.

The  theoretical models that have been proposed to study how
the chemical free energy can be converted into mechanical work
may be grouped into two classes.
One class of models\cite{magnasco94,keller00} introduce
a ``reaction (or chemical) coordinate'' $x$ as well as 
a ``mechanical coordinate'' $y$. 
The former coordinate describes how the chemical reaction proceeds
(the increase of  $x$ by, say, $a$ will correspond to the consumption of 
one fuel molecule due to the reaction) 
and the latter indicates how the engine translocates along a linear track 
(with a periodic structure of period $b$)
or how the rotor in the engine rotates ($b = 2\pi$).
The ``mechano-chemical coupling'' is expressed as a potential function
$U(x, y)$ that satisfies $U(x + a, y) = U(x, y) + \Delta G$ 
and $U(x, y + b) = U(x, y)$,
where $\Delta G < 0$ is the change in the free energy of the reservoir
containing fuel and product molecules due to the reaction of 
a single fuel molecule.
In this approach, the binding of a fuel molecule to and the unbinding of a
product molecule from the engine are not taken account explicitly.
As a result, the reservoir comes into play only through $\Delta G$,
which is unsatisfactory because the chemical potentials (or the densities)
of fuel and product molecules can affect independently the performance
of the engine, as one might expect or we shall see in this paper.

The other class of models deal with transitions between discrete chemical states
$j$ ($j = 1, 2, \dots, n$).
This class can be divided into two subclasses according to
whether the mechanical coordinate is discretized\cite{thomas01} 
or continuous.\cite{astumian96,wang98,parmeggiani99,kinoshita04}
In the former, 
the value of the mechanical coordinate is determined 
uniquely by the chemical state,
and hence a transition results in a translation of the engine (motor) or a rotation 
of the rotor.
In the latter, the engine in state $j$ undergoes Brownian motion under 
the influence of state-dependent potential $U_j(y)$ 
associated with the mechanical coordinate $y$,
and the transition from one state to another does no accompany the
translational or rotational motion of the engine at the instance of transition.
In this class of models, the binding of a reactant molecule to the engine, 
the chemical reaction, or the unbinding of a product molecule from the engine
trigger a transition between particular states.
However, the mechanism of energy transduction is not quite transparent
in the previous works on these models.

In the present work we take a combined approach:
a continuous reaction coordinate is used to describe 
the chemical reaction catalyzed by the engine, 
and discrete chemical states are introduced to
distinguish the state of the engine with a fuel or product molecule bound to it
from the state with no bound molecule,
while a continuous mechanical coordinate denotes the rotation of 
a rotor connected to the engine.
In constructing the model proposed here, we are inspired by the models for 
$\rm F_1$-ATPase, a biological rotary motor, 
by Wang and Oster\cite{wang98} and by Kinoshita et al.\cite{kinoshita04}
This protein machine is remarkable in that 
it works not only as a motor\cite{noji97,yasuda98}
(when ATP molecules are consumed) 
but also as a generator\cite{itoh04,rondelez05} 
(ATP  molecules are generated when it 
is forced to rotate in the reverse direction),  
just as a heat engine works as 
a heat pump when it is run in the reverse direction.
The minimal ingredients necessary to build a reversible molecular chemical engine
are extracted from their models\cite{wang98,kinoshita04} 
and an assumption on the chemical  reaction
is introduced to simplify the analysis.

Our model is simple enough to analytically work out extremely slow processes 
controlled externally,
which are analogous to the Carnot cycle of a heat engine.
We will call such a process a \textit{pseudo-static} process 
instead of a \textit{quasi-static} process,
because it turns out that such a slow process is not always reversible in 
the thermodynamic sense.
It will be shown that the average work delivered by the  engine
is identical to the free energy ($-\Delta G > 0$) released by 
the chemical reaction if the process is reversible,
and it is less than $|\Delta G|$ if the process is irreversible;
this result is consistent with the laws of thermodynamics.
However, it is found that in a single cycle of operation,
work larger than the maximum work imposed by thermodynamics
can be obtained by chance if the process is irreversible.

In the present model it is the torque resulting from the interaction between
the engine and a bound  (fuel or product) molecule
that contributes to the net mechanical work delivered in a cycle 
of pseudo-static process,
and this torque is independent of the chemical potentials
of fuel and product molecules in the reservoir (i.e., independent of $\Delta G$).
It will become clear that the chemical potentials determine the 
``timings'' of the binding and unbinding of fuel and product molecules,
thereby controlling the interval over which the torque is in effect.
Thus the amount of work delivered by the engine is affected by
the densities of fuel and product molecules in the reservoir.

\section{Model}
\label{sec:model}

We wish to construct a simple model for a molecular chemical engine
that is supposed to utilize a hypothetical chemical reaction
\begin{equation}
\label{eq:ab}
	A \rightleftarrows B.
\end{equation}
The setting we assume for the engine to work is as follows.
The engine is coupled to a rotor, 
whose rotational angle $\theta$ can be controlled
by an external agent. 
The engine exchanges molecules $A$ and $B$ with 
a reservoir that contains a large number of these molecules.
The chemical potentials $\mu_A$ and $\mu_B$ of 
molecules $A$ and $B$, respectively, are assumed to
satisfy the condition
\begin{equation}
\label{eq:muab}
	\Delta \mu \equiv \mu_A - \mu_B > 0.
\end{equation}
The engine and the reservoir is in contact with a heat bath
of temperature $T$.
The chemical reaction (\ref{eq:ab}) undergoes only at
a ``binding site'' on the engine, to which a single molecule $A$ or $B$ 
can be bound, 
and the engine has only one binding site.

It will be said that the engine is operated \textit{forward} or \textit{backward}
depending on whether the angle $\theta$ of the rotor is increasing
or decreasing.
The engine is designed so that it cannot bind a molecule 
when $\theta = 0$ (mod $2\pi$).
A forward (backward) \textit{cycle} is a process in which $\theta$ 
is increased (decreased) from zero to $2\pi$ ($-2\pi$);
the engine is empty both at the start and the end of a cycle.
A forward cycle is expected to proceed as follows.
A ``fuel'' molecule $A$ is bound to the engine.
Then a ``forward reaction'' $A \to B$ takes place at the binding site.
Finally, a ``waste'' (product) molecule $B$ is released to the reservoir.
Meanwhile, the engine delivers work by exerting torque on the rotor.
If the cycle proceeds as expected, the free energy of the reservoir changes 
by $\Delta G = -\Delta \mu < 0$ and the engine returns to the original state
after the cycle.
If the laws of thermodynamics can be applied to this process,
the maximum work the engine can deliver 
in  principle will be $\Delta \mu$.\cite{callen85}

Now we describe in detail how our model is constructed.

\subsection{Ligands}
\label{sec:ligands}

Molecules $A$ and $B$ are thought of as two stable states
of a single molecule, which will be referred to as a \textit{ligand}, 
with different geometrical structures.
It is assumed that a continuous transformation of structure from $A$ to $B$ 
can be characterized by a single, dimensionless parameter $r$,
a ``reaction coordinate.''

The energy profile associated with the structural change 
of a ligand molecule $U(r)$ may be defined,
apart from an unimportant additive constant, by
\begin{equation}
\label{eq:udef}
	\exp\left[-\beta U(r)\right] = \mathrm{Tr\,}\exp\left[-\beta H_L(r)\right],
\end{equation}
where $\beta = 1/k_\text{B}T$ with $k_\text{B}$ being the Boltzmann constant
and $H_L(r)$ is the Hamiltonian for the vibrational and rotational modes of 
a ligand molecule under the constraint that the reaction coordinate $r$ is held fixed;
we assume that $r$ can be treated as a classical dynamical variable.
In the definition (\ref{eq:udef}) of  $U(r)$, we have assumed, for simplicity, 
that the reservoir is an ideal gas mixture of molecules $A$ and $B$.
If the reservoir is a dilute solution where $A$ and $B$ are solutes, 
the interaction of a ligand with solvent molecules should be taken into
account in defining $U(r)$.
Once $U(r)$ is defined appropriately, the discussions presented below
will be valid irrespective of whether the reservoir is an ideal gas or a dilute solution.

In order for molecules $A$ and $B$ to be stable,
$U(r)$ must have two minima, say at $r = r_A$ and $r = r_B$, corresponding to
molecules $A$ and $B$.
The height of potential barrier separating these minima should be much larger
than the thermal energy $k_\text{B}T$ 
so that reaction (\ref{eq:ab}) does not occur spontaneously in the reservoir.

Let $n_A$ and $n_B$ be the densities (numbers of molecules per unit volume) of molecules 
$A$ and $B$, respectively, in the reservoir.
Then the chemical potentials of $A$ and $B$ are given by
\begin{equation}
\label{eq:muabn}
	\mu_\alpha = U_\alpha + k_\text{B}T\ln\frac{n_\alpha}{\eta_\alpha}
	\quad (\alpha = A, B),
\end{equation}
where $U_\alpha = U(r_\alpha)$, and
$\eta_\alpha$ is a quantity that has dimensions of density and is
proportional to a restricted partition function associated with the reaction coordinate: 
\begin{equation}
\label{eq:eta}
	\eta_\alpha \propto \int_{\mathcal{R}_\alpha}
	\exp\left\{-\beta[U(r) - U_\alpha]\right\}\,\d r
	\quad (\alpha = A, B).
\end{equation}
Here, $\mathcal{R}_\alpha$ is a small interval around $r = r_\alpha$ on the $r$ axis;
if $U(r)$ can be approximated by $U(r) = U_\alpha + \frac{1}{2}K_\alpha(r - r_\alpha)^2$
with a positive constant $K_\alpha$ in the vicinity of $r_\alpha$, 
we will have $\eta_\alpha \propto (2\pi\beta/K_\alpha)^{1/2}$.

\subsection{Engine with no ligand}
As already explained, the engine is coupled to a rotor
(through an arm-like structure, for example).
We imagine that the engine is a macromolecule like a biological molecular motor
and assume that its structure (conformation) changes with
the rotation of the rotor.
In this subsection we consider the case that the engine carries no ligand molecule.

Let $H_0(\theta)$ be the Hamiltonian of the engine with no ligand
under the condition that the rotation angle of the rotor remains fixed to be $\theta$.
The potential (free energy) $V_0(\theta)$ of the engine under this condition is
defined by
\begin{equation}
\label{eq:v0def}
	\exp\left[-\beta V_0(\theta)\right] 
	= \mathrm{Tr\,}\exp\left[-\beta H_0(\theta)\right].
\end{equation}
It is assumed that the engine is operated ($\theta$ is varied by the external agent)
so slowly  that it is in thermal equilibrium at any instant.
Then the torque $\tau_0(\theta)$  exerted by the engine on the rotor
is given as the statistical average of $-\partial H_0(\theta)/\partial \theta$, 
which implies
\begin{equation}
\label{eq:tau0}
	\tau_0(\theta) = -\frac{\d V_0(\theta)}{\d \theta}.
\end{equation}
Obviously, $V_0(\theta)$ and $\tau_0(\theta)$ possess the  following periodicity:
\begin{equation}
\label{eq:period0}
	V_0(\theta + 2\pi) = V_0(\theta), \quad
	\tau_0(\theta + 2\pi) = \tau_0(\theta).
\end{equation}

\begin{figure}[t]
	\begin{center}
		\includegraphics[width=7cm, clip]{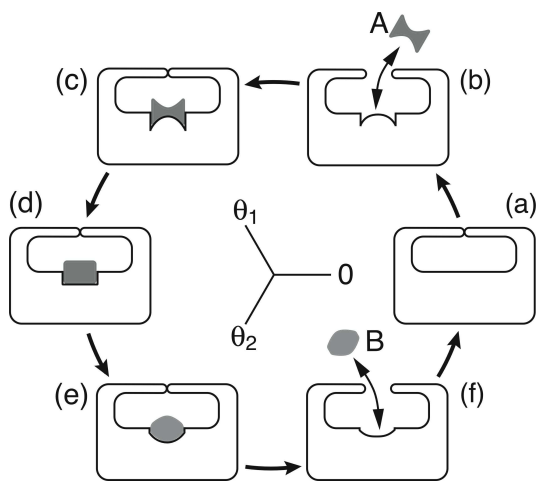}
	\end{center}
	\caption{
	The conformational change of the engine upon the rotation of the rotor
	is schematically shown. 
	The bold arrows indicate the direction of a forward operation.
	(a) $\theta \sim 0$, (b) $0 < \theta < \theta_1$, (c) $\theta \sim \theta_1$,
	(d) $\theta_1 < \theta < \theta_2$, (e) $\theta \sim \theta_2$, 
	(f) $\theta_2 < \theta < 2\pi$.
	}
	\label{fig:engine}
\end{figure}

In order that the engine works properly and can be analyzed without much difficulty
we require the engine to possess the following properties associated with its
conformational change.
Let $\theta_1$ and $\theta_2$ be some constants satisfying $0 < \theta_1 < \theta_2 < 2\pi$.
When the angle $\theta$ of the rotor is in the interval $0 < \theta < \theta_1$,
only molecule $A$ can be attached to or detached from the binding site of the engine;
if $\theta_1 < \theta < \theta_2$, no ligand can be attached or detached;
and when $\theta_2 < \theta < 2\pi$, only molecule $B$ can be attached or detached.
This can be made possible if, for example, the engine has a window-like structure
and a ligand can reach the binding site only by going through the opening of the window
(see Fig.~\ref{fig:engine}):
when $0 < \theta < \theta_1$, the opening of the window has a shape such that 
a molecule $A$ can go through it but a molecule $B$ cannot;
the window is closed if $\theta_1 < \theta < \theta_2$ or $\theta \sim 0$;
and when $\theta_2 < \theta < 2\pi$, the shape of the opening allows a molecule $B$ 
but not a molecule $A$ to  go through it.

\subsection{Engine with a ligand}
Now we consider the case that a ligand molecule is bound to the engine.
Let $H_\mathrm{int}(\theta, r)$ be the interaction between the engine and the ligand
under the constraint that the angle $\theta$ of the rotor and the  reaction coordinate $r$
of the ligand are held fixed.
We define potential $W(\theta, r)$ by
\begin{align}
\label{eq:wdef}
	&\exp\left\{-\beta[V_0(\theta) + W(\theta, r)]\right\}
	\nonumber\\
	&\quad{}= \mathrm{Tr\,}
	\exp\left\{-\beta[H_0(\theta) + H_L(r) + H_\mathrm{int}(\theta, r)]\right\},
\end{align}
where $V_0(\theta)$ has been defined in eq.~(\ref{eq:v0def}).
Roughly speaking, $W$ is the sum of the ligand energy and the engine-ligand interaction.

\begin{figure}[t]
	\begin{center}
		\includegraphics[width=7cm, clip]{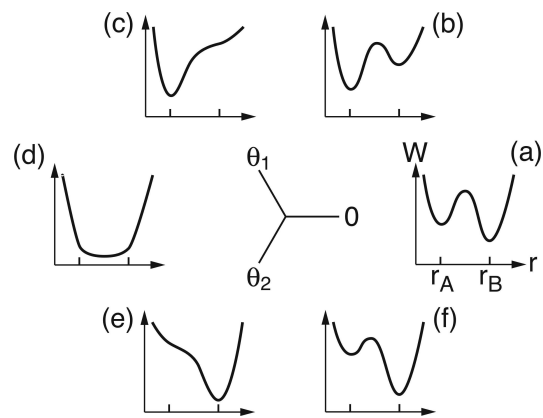}
	\end{center}
	\caption{
	Potential $W(\theta, r)$ defined by eq.~(\ref{eq:wdef}) as a function of $r$ for
	fixed values of $\theta$. 
	(a) $\theta \sim 0$, (b) $0 < \theta < \theta_1$, (c) $\theta \sim \theta_1$,
	(d) $\theta_1 < \theta < \theta_2$, (e) $\theta \sim \theta_2$, 
	(f) $\theta_2 < \theta < 2\pi$.
	}
	\label{fig:potw}
\end{figure}

The performance of the engine depends crucially on potential $W(\theta, r)$.
Let us describe the desired dependence of $W(\theta,  r)$  on $r$ for fixed values of $\theta$
(see Fig.~\ref{fig:potw}).
If the angle of the rotor is in interval $0 < \theta < \theta_1$,
$W$ should have a deep well around $r = r_A$ so that a molecule $A$ sitting at
the binding site stays there as molecule $A$ without being converted to molecule $B$;
it may have another minimum around $r = r_B$ separated by a large barrier.
Similarly, we require that $W$ should have a deep well around $r = r_B$ 
if $\theta_2 < \theta < 2\pi$,  so that molecule $B$ is stable at the binding site.
A ligand bound to the engine is supposed to undergo forward reaction $A \to B$ 
if $\theta$ is varied from $\theta_1$ to $\theta_2$
or backward reaction $B \to A$ if $\theta$ is varied from $\theta_2$ to $\theta_1$.
In this regard 
it is demanded that $W$ has only one minimum around $r = r_A$ for $\theta \sim \theta_1$, 
and only one minimum  around $r = r_B$ for $\theta \sim \theta_2$.
For $\theta_1 < \theta < \theta_2$, 
$W$ may have a single minimum at some $r$ between $r_A$ and $r_B$, 
and its location moves continuously as $\theta$ is varied between $\theta_1$ and $\theta_2$;
or $W$ may have a secondary minimum slightly above the primary one 
separated by a barrier whose hight is much less than $k_\mathrm{B}T$.

For the sake of simplicity the relaxation of the reaction coordinate $r$ is assumed to
occur quickly enough that this degree of freedom is always in thermal equilibrium
during an operation of the engine.
Then we can think of an ``effective interaction'' $\epsilon(\theta)$ defined by
\begin{equation}
\label{eq:epsdef}
	\exp\left[-\beta\epsilon(\theta)\right] 
	= \int_{\mathcal{R}(\theta)}\exp\left[-\beta W(\theta, r)\right]\,\d r,
\end{equation}
where the range $\mathcal{R}$ of integral depends on $\theta$ as follows:
$\mathcal{R}$ is the entire range of $r$ for $\theta_1 < \theta < \theta_2$,
$\mathcal{R} = \mathcal{R}_A$ [small interval around $r = r_A$ introduced in eq.~(\ref{eq:eta})]
for $0 < \theta < \theta_1$,
and $\mathcal{R} = \mathcal{R}_B$ for $\theta_2 < \theta < 2\pi$.
The reason why the integral range is restricted to $\mathcal{R}_A$ for $0 < \theta < \theta_1$ is
that $W$ as a function of $r$ may have a valley around $r = r_B$ as mentioned above
but this valley cannot be visited due to the large barrier separating it from the valley
around $r = r_A$.
For the similar reason the integral range is restricted to $\mathcal{R}_B$ for 
$\theta_2 < \theta < 2\pi$.
Because of these restrictions on the integral range in eq.~(\ref{eq:epsdef}),
$\epsilon(\theta)$ is discontinuous at $\theta = 0$ (mod $2\pi$) in general,
thought it is periodic, $\epsilon(\theta + 2\pi) = \epsilon(\theta)$.
It is not difficult to see that the engine carrying a ligand exerts torque
\begin{equation}
\label{eq:tau1}
	\tau_1(\theta) = - \frac{\d V_0(\theta)}{\d\theta} - \frac{\d\epsilon(\theta)}{\d\theta}
\end{equation}
on the rotor when its rotational angle is $\theta$;
it is the statistical average of 
$
	-\partial[H_0(\theta) + H_\text{int}(\theta, r)]/\partial \theta.
$

\subsection{Binding and unbinding of a ligand}

Suppose that the rotational angle $\theta$ of the rotor is in range $0 < \theta < \theta_1$.
Then the binding site of the engine can be empty or occupied by a molecule $A$.
If the binding-unbinding process is in equilibrium, 
the probability $p_A(\theta)$ that a molecule $A$ is bound to the engine is given by
\begin{equation}
\label{eq:pa0}
	p_A(\theta) = \frac{Z_1(\theta)\exp(\beta \mu_A)}
		{Z_0(\theta) + Z_1(\theta)\exp(\beta \mu_A)},
\end{equation}
where $Z_0$ and $Z_1$ are, respectively, the partition functions of the engine
without and with a ligand.
From the definitions of $V_0(\theta)$ and $\epsilon(\theta)$ 
we have
\begin{align}
\label{eq:z0}
	Z_0(\theta) &= \exp\left[-\beta V_0(\theta)\right], 
	\\
\label{eq:z1}
	Z_1(\theta) &= \exp\left\{-\beta[V_0(\theta) + \epsilon(\theta)]\right\},
\end{align}
and therefore eq.~(\ref{eq:pa0}) can be expressed as
\begin{equation}
\label{eq:pa}
	p_A(\theta) = \frac{1}{\exp\left\{\beta[\epsilon(\theta) - \mu_A]\right\} + 1}
	\quad
	(0 < \theta < \theta_1).
\end{equation}
Similarly, we have
\begin{equation}
\label{eq:pb}
	p_B(\theta) = \frac{1}{\exp\left\{\beta[\epsilon(\theta) - \mu_B]\right\} + 1}
	\quad
	(\theta_2 < \theta < 2\pi)
\end{equation}
for the probability that a molecule $B$ is bound to the engine when
$\theta$ is in interval $\theta_2 < \theta < 2\pi$.

We have assumed that the engine is designed so that a ligand cannot be bound
when $\theta = 0$ (mod $2\pi$).
Considering eqs.~(\ref{eq:pa}) and (\ref{eq:pb}),
this requirement implies that
\begin{align}
\label{eq:e0cond}
	&\exp\left\{-\beta[\epsilon(+0) - \mu_A]\right\} = 0, 
	\\
\label{eq:e2picond}
	&\exp\left\{-\beta[\epsilon(2\pi - 0) - \mu_B]\right\} = 0.
\end{align}
(Remember that in general $\epsilon(\theta)$ is discontinuous at $\theta = 0$,
i.e., $\epsilon(0+) \not= \epsilon(2\pi - 0)$.)
In reality these conditions cannot be satisfied exactly, 
because $\epsilon$ cannot be infinite.
We consider eqs.~(\ref{eq:e0cond}) and (\ref{eq:e2picond}) as mathematical idealization
for a situation in which the left-hand sides of  these equations
are extremely small and can be treated as zero practically.

\begin{figure}[t]
	\begin{center}
		\includegraphics[width=7cm, clip]{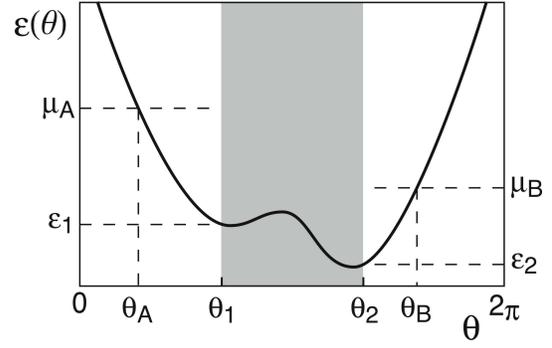}
	\end{center}
	\caption{
	An example of effective interaction $\epsilon(\theta)$ between the engine 
	and a bound ligand defined by eq.~(\ref{eq:epsdef}).
	Binding and unbinding transitions of a molecule $A$ occurs for $0 < \theta < \theta_1$.
	Chemical reaction $A \rightleftarrows B$ undergoes if a ligand is bound for 
	$\theta_1 < \theta < \theta_2$; in this interval both binding and unbinding of
	a ligand is prohibited.
	Binding and unbinding transitions of a molecule $B$ take place for 
	$\theta_2 < \theta < 2\pi$.
	}
	\label{fig:eps}
\end{figure}

Figure \ref{fig:eps} shows an example of ``effective interaction'' $\epsilon(\theta)$
between the engine and a ligand that satisfies conditions (\ref{eq:e0cond}) and
(\ref{eq:e2picond}) and certain properties we are going to discuss.
When the engine is operated forward, 
it is preferable that probability $p_A(\theta)$ increases as $\theta$ is increased
from zero to $\theta_1$,
while $p_B(\theta)$  decreases as $\theta$ is varied from $\theta_2$ to $2\pi$.
Hence, we assume that $\epsilon(\theta)$ decreases with increasing $\theta$ 
for $0 < \theta < \theta_1$ and increases with $\theta$ for $\theta_2 < \theta < 2\pi$.
The dependence of $\epsilon$ on $\theta$ for $\theta_1 < \theta < \theta_2$ 
can be rather arbitrary,
because binding or unbinding of a ligand is prohibited in this interval.

If the chemical potential $\mu_A$ of molecule $A$ lies well above 
$\epsilon_1 \equiv \epsilon(\theta_1)$,
the engine carries a molecule $A$ with large probability when $\theta$ reaches $\theta_1$
in a forward operation.
Similarly, the engine carries a molecule $B$ with large probability 
when $\theta$ is decreased to $\theta_2$
in a backward operation if $\mu_B$ lies well above $\epsilon_2 \equiv \epsilon(\theta_2)$.
These conditions are favorable for good performance of the engine.
We shall not assume, however, that these are always satisfied, 
because we are interested in how the engine may work under poor conditions
as well as ideal ones.

\section{\label{sec:operation}Pseudo-Static Operations of the Engine}

Let us analyze the work delivered by the engine when it is 
operated slowly enough that the binding and unbinding of 
ligands is in equilibrium at almost any instance  of the operation.
We call this process a \textit{pseudo-static} process to distinguish
it from a \textit{quasi-static} process;
the latter is preserved for a reversible process,
while the former may be reversible or irreversible as we shall  see.
In a pseudo-static process the probability that the engine carries a ligand 
is given by eq.~(\ref{eq:pa}) for $0 < \theta <  \theta_1$
and by eq.~(\ref{eq:pb}) for $\theta_2 < \theta < 2\pi$,
and the torque exerted on the rotor by the engine is given by
\begin{equation}
\label{eq:taua}
	\tau(\theta) = [1 - p_A(\theta)]\tau_0(\theta) + p_A(\theta)\tau_1(\theta)
\end{equation}
for $0 < \theta < \theta_1$ and by
\begin{equation}
\label{eq:taub}
	\tau(\theta) = [1 - p_B(\theta)]\tau_0(\theta) + p_B(\theta)\tau_1(\theta)
\end{equation}
for $\theta_2 < \theta < 2\pi$, 
where $\tau_0$ and $\tau_1$ are defined in eqs.~(\ref{eq:tau0}) and
(\ref{eq:tau1}), respectively.
These expressions for the torque can be rewritten as
\begin{equation}
\label{eq:tauab}
	\tau(\theta) = -\frac{\d}{\d \theta}
		\left\{V_0(\theta) + \frac{1}{\beta}\ln[1 - p_\alpha(\theta)]\right\},
\end{equation}
where $\alpha = A$ for $0 < \theta < \theta_1$ and $\alpha = B$ for 
$\theta_2 < \theta < 2\pi$.
Note that the quantity in the braces in eq.~(\ref{eq:tauab}) is identical to
the grand potential (grand-canonical free energy) of the engine.\cite{shibata00}
For $\theta_1 < \theta < \theta_2$ the  torque is given by 
$\tau(\theta) = \tau_0(\theta)$ or $\tau(\theta) = \tau_1(\theta)$ depending on 
whether the engine carries no ligand or one.

\subsection{\label{sec:forward}Forward operation}
In a forward, pseudo-static operation of the engine the external agent is supposed to
control the angle $\theta$ of the rotor to increase extremely slowly.
The work delivered by the engine in one cycle of forward operation
is given by
\begin{equation}
\label{eq:wfw}
	W_\text{fw} = \int_0^{2\pi}\tau(\theta)\,\text{d}\theta,
\end{equation}
where $\tau(\theta)$ is the torque exerted by the engine.

We divide one cycle of a forward operation into three \textit{phases}:
the period in which $0 < \theta < \theta_1$ will be called ``phase $A$,'' 
the period of $\theta_1 < \theta < \theta_2$ ``phase $C$,''
and the last one ($\theta_2 < \theta < 2\pi$) ``phase $B$.''
In phase $A$ the torque exerted by the engine is given by eq.~(\ref{eq:tauab}) 
with $\alpha = A$, and therefore the work delivered in this phase is found to be
\begin{equation}
\label{eq:wfwa}
	W_\text{fw}^A = V_0(0) - V_0(\theta_1) - \frac{1}{\beta}\ln(1 - p_1),
\end{equation}
where probability $p_1$ is defined by
\begin{equation}
\label{eq:p1def}
	p_1  = p_A(\theta_1).
\end{equation}
Condition (\ref{eq:e0cond}) has been used in deriving eq.~(\ref{eq:wfwa}). 
The work $W_\text{fw}^B$ obtained in phase $B$ can be worked out
similarly.
Using expression (\ref{eq:tauab}) with $\alpha = B$ and
condition (\ref{eq:e2picond}), we get
\begin{equation}
\label{eq:wfwb}
	W_\text{fw}^B = V_0(\theta_2)  - V_0(2\pi) + \frac{1}{\beta}\ln(1 - p_2),
\end{equation}
where probability $p_2$ is defined by
\begin{equation}
\label{eq:p2def}
	p_2 = p_B(\theta_2).
\end{equation}
In phase $C$ there are two alternatives:
if the engine carries no ligand, the work delivered in this phase is
\begin{equation}
\label{eq:wfwc0}
	W_\text{fw}^{C0} = V_0(\theta_1) - V_0(\theta_2)
\end{equation}
since the torque is given by $\tau_0(\theta)$;
and if a ligand is bound to the engine, the work is given by
\begin{equation}
\label{eq:wfwc1}
	W_\text{fw}^{C1} = V_0(\theta_1) - V_0(\theta_2)
		+ \epsilon(\theta_1) - \epsilon(\theta_2)
\end{equation}
since the torque is $\tau_1(\theta)$.
Summing up the contributions from the three phases, we obtain
\begin{equation}
\label{eq:wfw0}
	W_\text{fw}^0 = \frac{1}{\beta}\ln\frac{1 - p_2}{1 - p_1}
\end{equation}
for the work delivered in a cycle in the case that the engine
carries no ligand in phase $C$.
If the engine carries a ligand in phase $C$,
the work in one cycle is given by
\begin{equation}
\label{eq:wfw1}
	W_\text{fw}^1 = \Delta \mu + \frac{1}{\beta}\ln\frac{p_2}{p_1}.
\end{equation}

Let us examine the results for $W_\text{fw}^0$ and $W_\text{fw}^1$ 
obtained above.
It is noted that depending on the design of the engine and/or 
the preparation of the reservoir any of the cases (i) $p_1 = p_2$, 
(ii) $p_1 < p_2$, and (iii) $p_1 > p_2$ can be realized.
In the first case we have
\begin{equation}
\label{eq:wcase1}
	W_\text{fw}^0 = 0, \quad W_\text{fw}^1 = \Delta \mu.
\end{equation}
This consequence is in accordance with our naive expectation that
the engine delivers no work if it does not catalyze the chemical reaction,
and the free energy $\Delta \mu$ released in reaction $A \to B$
is converted into mechanical work if the engine catalyzes the reaction.
In the case of $p_1 < p_2$ one finds
\begin{equation}
\label{eq:wcase2}
	W_\text{fw}^0 < 0, \quad W_\text{fw}^1 > \Delta \mu.
\end{equation}
Note that the second inequality tells that \textit{work greater than the 
the maximum work dictated by the second law of thermodynamics
can be extracted in one cycle of operation}.
Apparent violation of the second law may occur also in the case of $p_1 > p_2$,
in which we observe
\begin{equation}
\label{eq:wcase3}
	W_\text{fw}^0 > 0, \quad W_\text{fw}^1 < \Delta \mu.
\end{equation}
The first inequality implies that the heat extracted from a single heat bath
can be converted into mechanical work without leaving any changes
in the engine and the reservoir of ligand molecules, 
which appears to be in contradiction to the second law.

The fact that the engine can deliver work larger than what is expected 
from thermodynamics  in one cycle of operation does not necessarily 
mean inadequacy of the present model.
This is because such an event occurs stochastically and in an uncontrollable manner.
In the case of $p_1 < p_2$, for example, the work delivered 
may be  larger than the maximum work limited by thermodynamics
($W_\text{fw}^1 > \Delta \mu$) in one cycle 
but it may be less than the maximum work ($W_\text{fw}^0 < 0$) in another.
Therefore, the appropriate quantity to be worked out should be 
the average work per cycle in a forward operation carried out 
over many cycles.
In a forward operation 
the probability that a ligand is bound in phase $C$ is given by $p_1$.
Hence, the average work delivered per cycle is 
\begin{equation}
\label{eq:wfwav}
	W_\text{fw} = (1 - p_1)W_\text{fw}^0 + p_1W_\text{fw}^1
	= p_1\Delta\mu - \frac{1}{\beta}f(p_1, p_2),
\end{equation}
where function $f(x,y)$ is defined by
\begin{equation}
\label{eq:fxy}
	f(x,y) = (1 - x)\ln\frac{1 - x}{1 - y} + x\ln\frac{x}{y}.
\end{equation}
It is easy to see that $f(x, y) \ge 0$ for $0 \le x \le 1$ and $0 \le y \le 1$,
where the equality holds if $x = y$.
Thus we find
\begin{equation}
\label{eq:wfwineq}
	W_\text{fw} \le p_1\Delta\mu,
\end{equation}
where the equality holds if $p_1 = p_2$.
If a ligand is bound in phase $C$ of a forward cycle (with probability $p_1$),
a molecule $A$ in the reservoir will have been transformed into a molecule $B$
when the cycle is completed.
This means that the average decrease in the free energy of the reservoir per cycle
is given by $p_1\Delta \mu$, 
which is identical to the right-hand side of eq.~(\ref{eq:wfwineq}).
Consequently, it turns out that inequality (\ref{eq:wfwineq}) is 
\textit{consistent with the maximum work theorem of thermodynamics},
and the maximum work is obtained if $p_1 = p_2$.

\subsection{\label{sec:backward}Backward operation}
The angle $\theta$ of the rotor is decreased extremely slowly
by the external agent in a pseudo-static, backward operation of the engine.
A cycle starts with phase $B$ in which $\theta$ is decreased from $2\pi$ to $\theta_2$,
and then phase $C$ ($\theta_2 > \theta > \theta_1$) and 
phase $A$ ($\theta_1 > \theta > 0$) follow.
The work delivered by the engine in each phase is obtained from the corresponding work
in a forward operation by changing its sign.
Consequently, the work in  a cycle is given by
\begin{equation}
\label{eq:wbw0}
	W_\text{bw}^0 = \frac{1}{\beta}\ln\frac{1 - p_1}{1 - p_2}
\end{equation}
if a ligand is not bound in phase $C$, and by
\begin{equation}
\label{eq:wbw1}
	W_\text{bw}^1 = -\Delta\mu + \frac{1}{\beta}\ln\frac{p_1}{p_2}
\end{equation}
if a ligand is bound in phase $C$.

Since a ligand is bound with probability $p_2$ in phase $C$ of a backward operation, 
the average work delivered by the engine per cycle turns out to be
\begin{equation}
\label{eq:wbwav}
	W_\text{bw} = (1 - p_2)W_\text{bw}^0 + p_2W_\text{bw}^1
	= -p_2\Delta\mu - \frac{1}{\beta}f(p_2, p_1),
\end{equation}
where function $f(x,y)$ has been defined in eq.~(\ref{eq:fxy}).
Since $f(x, y) \ge 0$ as remarked earlier,  it is concluded from eq.~(\ref{eq:wbwav}) that
\begin{equation}
\label{eq:wbwineq}
	-W_\text{bw} \ge p_2\Delta\mu,
\end{equation}
where the equality holds if $p_1 = p_2$.

When the engine is operated backwards, 
it works as a \textit{generator} in the sense that fuel molecules $A$ 
are generated from waste molecules $B$.
As a result, the free energy of the reservoir is increased.
The right-hand side of inequality (\ref{eq:wbwineq}) represents the average increase in 
the free energy of the reservoir per cycle, while 
the left-hand side is the work provided by the external agent.
Therefore, this inequality tells that the minimum work necessary for converting 
molecules $B$ into $A$ is equal to the change in the free energy of the reservoir,
which is \textit{consistent with the minimum work theorem of thermodynamics}.

\subsection{\label{sec:reversibility}Reversibility}
It is evident that a single cycle of a forward or backward operation discussed above 
cannot be reversible unless $p_1 = p_2 = 1$.
Suppose, for example, that a ligand is bound in phase $C$ of a forward cycle
and the engine is operated backwards after the cycle is completed.
The engine may or may not carry a ligand in phase $C$ of the backward cycle,
indicating that individual cycles are not reversible.

A cycle may be reversible, however, in the following statistical sense.
Considering the average over many cycles, 
the probability that the engine carries a ligand in phase $C$ is $p_1$ in a
forward cycle  and $p_2$ in a backward cycle.
Therefore, if the condition $p_1 = p_2$ is satisfied,
the statistical properties of the whole system 
(including the engine, the reservoir of ligands, the heat bath and the external agent)
at a given value of $\theta$ in a forward cycle is exactly the same as
that in a backward cycle.
This condition of reversibility coincides with the condition for equalities 
in eqs.~(\ref{eq:wfwineq}) and (\ref{eq:wbwineq}) to hold.
This observation is consistent with thermodynamics:
\textit{the maximum work obtained or the minimum work needed to provide in a process 
is achieved when the process is reversible}.

It should be remarked that, even in the statistical sense,
pseudo-static operations of the engine discussed above
are not reversible  if $p_1 \not= p_2$.
The irreversibility arises when the phase of operation changes from $C$ to $B$
in a forward process, for example:
The probability that a ligand is bound to the engine is $p_1$ for $\theta = \theta_2 - 0$
(phase $C$), and right after $\theta$ is increased across $\theta_2$
it must be relaxed irreversibly towards $p_2$, the probability in equilibrium for  
$\theta = \theta_2 + 0$ (phase $B$).
In  a backward cycle the irreversibility comes out upon passing from phase $C$ to phase $A$.
This kind of irreversibility cannot  be eliminated no matter how slowly $\theta$ is varied.
The same kind of irreversibility is discussed in ref.~\citen{sekimoto04},
and a similar irreversibility is pointed out\cite{sato02} to result from 
the contact of a small system with a heat bath.

\section{\label{sec:mechanism}Mechanism of Energy Transduction}

The present model of a molecular chemical engine is simple enough
to find out how the engine converts the free energy (\ref{eq:muab})
released by the chemical reaction into mechanical work.
The torque exerted by the engine has contributions from
the conformational free energy $V_0(\theta)$ of the engine without a ligand
and from the effective interaction $\epsilon(\theta)$ between the engine
and a bound ligand.
It is evident that the $\epsilon$ is responsible for net mechanical work
in one cycle of operation, because the contribution from $V_0$ cancels out
over a cycle due to its periodicity (\ref{eq:period0}).
However, the interaction $\epsilon(\theta)$ appears to have nothing to do
with the chemical potentials $\mu_A$ and $\mu_B$
of reactant and product molecules, 
which define the decrease $\Delta \mu$ in the free energy of the reservoir.
Then, how the work delivered by the engine may depend on $\mu_A$ and
$\mu_B$?
We will see that \textit{the chemical potentials affect the ``timings'' 
of switching ``on'' and ``off'' the interaction $\epsilon(\theta)$, and
that the amount of work delivered by the engine during a cycle
depends on these timings}.

We first consider an idealized situation in which
$\mu_A$ lies well above $\epsilon_1 = \epsilon(\theta_1)$ and 
$\mu_B$ well above $\epsilon_2 = \epsilon(\theta_2)$ (see Fig.~\ref{fig:eps}), 
and $k_\text{B}T$ is much smaller than $\mu_A - \epsilon_1$
and $\mu_B - \epsilon_2$.
In this circumstance
we have $p_1 \simeq 1$ and $p_2 \simeq 1$,
and observe that $p_A(\theta)$ increases from zero to one
in a narrow region around $\theta_A$ defined by $\epsilon(\theta_A) = \mu_A$
as $\theta$ is increased from 0 to $\theta_1$, according to eq.~(\ref{eq:pa}).
Similarly, $p_B(\theta)$ decreases from one to zero around angle $\theta_B$
defined by $\epsilon(\theta_B) = \mu_B$ as
$\theta$ is increased from $\theta_2$ to $2\pi$.
This means that the engine catches a molecule $A$ when $\theta$ reaches around  $\theta_A$
in a forward cycle, and keeps carrying the ligand afterwards until
$\theta$ reaches $\theta_B$, 
where the ligand is released as a molecule $B$.
Hence, the interaction $\epsilon(\theta)$ is switched on around $\theta = \theta_A$
and switched off around $\theta = \theta_B$ in this situation.
Consequently, the work delivered by the engine is estimated to be
\begin{equation}
\label{eq:wideal}
	W \simeq \mu_A - \mu_B, 
\end{equation}
because the torque due to $\epsilon(\theta)$
is given by $-\d \epsilon/\d \theta$ [the second term in eq.~(\ref{eq:tau1})].
This result is consistent with the rigorous one (\ref{eq:wfw1}) with
$p_1 \simeq 1$ and $p_2 \simeq 1$.
In this way, we see that the timings of binding and unbinding of a ligand molecule
determined by $\mu_A$ and $\mu_B$ control the work delivered by the engine:
if $\mu_A$ is increased, for example, then $\theta_A$ decreases and hence
the range of $\theta$ over which $\epsilon(\theta)$ is on increases; 
this results in the increase in the work delivered by the engine, 
since $\epsilon(\theta)$ in this interval yields positive torque.

In the above argument, 
the temperature is assumed to be low enough
that the switching of the interaction $\epsilon(\theta)$ occur sharply 
around $\theta = \theta_A$ and $\theta = \theta_B$.
If the temperature is not so low the switching occur gradually:
for example, binding and unbinding of molecule $A$ are repeated many times
over a certain range of $\theta$ around $\theta_A$.
However, we can say that $\epsilon(\theta)$ is switched on at $\theta = \theta_A$
\textit{on average},
because probability $p_A$ given by eq.~(\ref{eq:pa}) regarded as a function of $\epsilon(\theta)$ 
is ``antisymmetric'' about $\epsilon = \mu_A$:
transformation $\epsilon - \mu_A \to \mu_A - \epsilon$ results in $p_A \to  1 - p_A$. 
Similarly, it can be said that $\epsilon(\theta)$ is switched off at $\theta_B$ on average
as long as $p_2 \simeq 1$.
Hence, the argument leading to eq.~(\ref{eq:wideal}) is still valid even in the case of
gradual switching, if the timings of switching are understood as the averaged ones
discussed here.

In the idealized situation considered above,
the engine works with high efficiency for arbitrary choices of $\mu_A$ and $\mu_B$
as far as conditions $p_1 \simeq 1$ and $p_2 \simeq 1$ are satisfied:
the decrease in the free energy (\ref{eq:muab}) of the reservoir is almost
perfectly converted into mechanical work ($W \simeq \Delta \mu$).
If either of these conditions is not satisfied, the efficiency is not always good.
For example, suppose that $\mu_B$ lies well below $\epsilon_2$ ($p_2 \simeq 0$),
and $\mu_A$ well above $\epsilon_1$ ($p_1 \simeq 1$).
In a forward cycle under this extreme circumstance, 
the engine catch a molecule $A$ at $\theta = \theta_A$ on average (since $p_1 \simeq 1$), 
and releases a molecule $B$ immediately after the process enters phase $B$ (since $p_2 \simeq 0$).
Therefore, potential $\epsilon(\theta)$ is on for $\theta_A < \theta < \theta_2$
on average,
and the work delivered in a cycle is estimated as 
\begin{equation}
\label{eq:workex1}
	W \simeq \mu_A  - \epsilon_2,
\end{equation}
which is independent of $\mu_B$ and  less than $\Delta \mu$.
This result is consistent with the rigorous one given by (\ref{eq:wfw1}) with
$p_1 \simeq 1$ and $p_2 \simeq \exp[\beta(\mu_B - \epsilon_2)]$,
where the latter expression is obtained from eq.~(\ref{eq:pb}) under the condition that $p_2 \ll 1$.
According to eq.~(\ref{eq:muabn}), 
$\Delta \mu$ can become large indefinitely
by letting the density of molecule $B$ in the reservoir become vanishingly small
($\mu_B \to -\infty$).
However, the work remains unaltered and the efficiency $W/\Delta\mu$ goes to  zero 
in this limit.
A similar, qualitative argument is given in Ref.~\citen{kinoshita04}.

Another extreme case of interest may be the situation in which 
conditions
\begin{equation}
\label{eq:cond}
	\epsilon_1 > \mu_A > \mu_B > \epsilon_2,
\end{equation}
$p_1 \ll 1$ and $p_2 \simeq 1$ are satisfied.
Such a situation can be realized by lowering the density of fuel molecule in the reservoir.
In a forward cycle under this circumstance,
a fuel molecule $A$ can be bound to the engine only in the vicinity of $\theta = \theta_1$ 
(with very small probability).
If the engine carries a ligand when the cycle proceeds to phase $C$, 
a forward reaction $A \to B$ takes place and a molecule $B$ will be released
around $\theta = \theta_B$ (since $p_2 \simeq 1$).
Therefore, the work delivered in this cycle will be
\begin{equation}
\label{eq:workex2}
	W \simeq \epsilon_1 - \mu_B,
\end{equation}
which is larger than $\Delta \mu = \mu_A - \mu_B$ according to
condition (\ref{eq:cond}).
This result is again consistent with the rigorous one given by (\ref{eq:wfw1}) with
$p_1 \ll 1$ and $p_2 \simeq 1$,
and provides an example in which work larger than the maximum work predicted by
thermodynamics can be obtained in a single cycle of forward operation
[the second inequality in eq.~(\ref{eq:wcase2})].

\section{\label{sec:conclusion}Concluding Remarks}

We have proposed a simple, idealized model for a molecular chemical engine,
and analyzed pseudo-static processes associated with it.
The work delivered by the engine in a forward operation (\S~\ref{sec:forward}) 
and the work necessary to operate the engine backwards (\S~\ref{sec:backward})
have been calculated analytically.
The statistical averages  of these works are found to be consistent with
the maximum/minimum work theorem of thermodynamics,
whereas the work for an individual cycle of operation is not necessarily so.
We have also clarified (\S~\ref{sec:mechanism}) how the densities of fuel and product molecules 
in the reservoir, which determine the free energy released 
by the chemical reaction,
can affect the work delivered by the engine.

In the present work, we have restricted our attention to 
pseudo-static processes controlled by an external agent.
It is possible to make the model engine constructed here
run autonomously (i.e., without the external agent)
and do work against a load torque applied to the rotor.
The angular velocity of the rotor,
the thermodynamic efficiency and other quantities 
characterizing the performance of the engine are under investigation.

The recent experiment on $\rm F_1$-ATPase,\cite{rondelez05}
a biological rotary motor, 
has clearly shown that it consumes three ATP molecules
per turn of the rotor (the $\gamma$ subunit) in a forward operation, 
while a forced, reversal rotation of the rotor produces about
2.3 ATP molecules per turn on average.
Although the experimental situation cannot be considered as
pseudo-static, it would be tempting to compare it with the
present analysis.
The experimental results mentioned above indicate that 
$p_1 \simeq 1$ and $p_2 \simeq 0.77$,
since $\rm F_1$-ATPase contains three engines called the $\beta$ subunits
and each engine is thought to be responsible for $120^\circ$ rotation of
the rotor.
In our model, probability $p_2$ depends on the  concentration of product molecules
in the reservoir.
In the case of $\rm F_1$-ATPase, 
the product molecules are adenosine diphosphate (ADP) and 
inorganic phosphate (Pi).
The physiological concentrations of ATP and Pi were chosen in the experiment.
If our simple model captures the essence of energy transduction 
in $\rm F_1$-ATPase, it is predicted that different choices of 
ADP and Pi concentrations will result in different values of $p_2$ 
(mechano-chemical coupling efficiency);
$p_2$ will increase if these concentrations are increased.

Finally, we mention a few theoretical works that have closely 
related with the present work.
A quasi-static process of a macroscopic chemical engine is  discussed
in ref.~\citen{shibata98}.
In refs.~\citen{sasa98,sekimoto04} models of molecular motors
driven by two reservers containing the same kind of particles with different densities;
the decrease in the total free energy of the reservoirs resulting from the 
transfer of particles from the high-density reservoir to the low-density one
through the motor is converted into mechanical work.
The Carnot cycle for a microscopic heat engine 
is analyzed in ref.~\citen{sekimoto00}.

\section*{Acknowledgements}
I would like to thank Professor E.~Muneyuki for a stimulating discussion, which encourage me to start the present work.
I also thank Y.~Ito for a number of discussions
on the present model in its infancy.
This work is supported in part by
the Grants-in-Aid for
Scientific Research in Priority Areas from the Ministry
of Education, Culture, Sports, Scinece and Technology.

\end{document}